\documentstyle[prl,aps,epsf,multicol]{revtex}

\begin{document}

\draft

\title{A Simple Model of Evolution with Variable System Size}

\author{Claus Wilke and Thomas Martinetz}

\address{Institut f\"ur Neuroinformatik\\Ruhr-Universit\"at Bochum}

\date{Submitted: ; Printed: \today}

\maketitle

\begin{abstract}
A simple model of biological extinction
with variable system size which exhibits a power-law
distribution of 
extinction event sizes is presented. The model is a generalization of a
model recently introduced by Newman (Proc. R. Soc. Lond. B{\bf 263},
  1605 (1996)). Both analytical and numerical
analysis show that the exponent of the power-law distribution depends
only marginally on the growth rate $g$ at which new species enter
the system and is equal to the one of the original model in the limit
$g\rightarrow\infty$. A critical growth rate $g_c$,
below which the system dies out, can be found. Under these model
assumptions stable 
ecosystems can only exist if the regrowth of species is sufficiently fast.

\end{abstract}



\begin{multicols}{2}
The fact that extinction events seem to be episodic on all scales, as
noted by Raup~\cite{Raup}, has
aroused much interest in the last few years. Throughout the history of
life on earth there have been many small
extinction events, but very big ones have happened only rarely. A
histogram of the 
frequency of extinction events of different sizes indicates a
power law distribution $p(s)=s^{-\tau}$, where $s$ denotes
the number of species that go extinct in one event and $p(s)$ denotes
the frequency of events of size~$s$.

There are two mechanisms to explain mass extinctions. On the one hand, it is 
argued that coevolution can drive large proportions of an ecosystem into 
extinction and produce extinction events on all scales. 
Ecosystems might drive themselves into a critical
state in which a small change (e.g. the mutation of a single species) can
trigger an ``avalanche'' that may span the whole system. For this kind
of dynamic Bak et al.~\cite{Bak1} have coined the name Self-Organized
Criticality. Several simple models of evolution exhibiting SOC have
been proposed, among them models by Kauffman and
Johnsen~\cite{Kauffman1}, Bak and  
Sneppen~\cite{Bak2}, Manrubia and Paczuski~\cite{Manrubia}. 

On the other
hand, it is argued that mass extinctions find  
their origin in external influences. That situation is modelled by
some recent work of Newman~\cite{Newman2}. He used a model belonging
to the new class of so called ``coherent noise'' models recently introduced
by Newman and Sneppen~\cite{Newman1}.
These models 
are clearly not SOC but they nevertheless show a power law distribution of
avalanche sizes. Newman compared his model
with the analysis of the fossil record performed by Raup. 
The exponent $\tau$ close to 2 that arises in this model is in 
good agreement with the fossil record. Thus Newman came to the
conclusion that there is no evidence for SOC as the  
major driving force for extinction.

It can be generally observed that the majority of the models for
biological evolution and extinction  
up to now considered work with a fixed number of species. This is a
major drawback since it is in clear
contrast with the biological reality. After a major extinction event, the
number of species in the ecosystem is significantly reduced, and the
process of regrowth of new species can take a long time. The fossil
record~\cite{Benton} shows that the process of 
growth of species is commonly interrupted by extinction events. 

To our knowledge,
models with variable system size have only been studied by Vandewalle and 
Auslool~\cite{Vandewalle} and by Head and Rodgers~\cite{Head}. But in
both cases the  
models do not explain the distribution of extinction events seen in the fossil 
record. The model of Vandewalle and Auslool is a tree model that grows 
infinitely, while the model of Head and Rodgers reaches a steady-state
in which no major extinctions occur. As far as we know, none of the
models with variable  
system size up to now considered can explain the distribution of 
extinction events seen in the fossil record.

But every mechanism proposed for the explanation of mass extinctions must
\begin{itemize}
 \item explain the distribution seen in the fossil record,
  \item face the fact that the number of species is not constant, but
    is reduced significantly after a major extinction event.
\end{itemize}
A priori it is not at all clear if a mechanism producing a certain 
distribution of extinction events will show the same distribution when the 
constraint of a fixed system size is released. Therefore it is very
important to  
study models with variable system size.

We 
propose here a generalization to the coherent noise model used by
Newman, where the 
refilling of the system is
done in finite time.
Newman's model is defined as follows. The system
consists of $N$ species, each possessing a threshold $x_i$ of tolerance
against stress, chosen from a probablity distribution $p_{\rm
  thresh}(x)$. At each time-step, a stress $\eta$ is generated at random with a
distribution $p_{\rm stress}(\eta)$, and all species with $x_i<\eta$ are
removed from the system and immediately replaced with new
ones. Furthermore, a small fraction $f$ of the species is chosen at
random and given new thresholds. That corresponds to a probability of
$f$ for every species to undergo spontaneous mutation. 

In our model the fraction of species with
$x_i<\eta$ is removed permanently from the system, but in every
time-step there is some growth of new species. 

Note that the generalized model, like the original one, does not include
explicitly interaction between species. There are two reasons to
justify this model assumption. Firstly, previous work\cite{Newman3}
has shown that 
the coherent noise dynamic is very strong and can dominate interaction
dynamic. Secondly, the investigation of a model without interaction, that
can reproduce important features of the fossil record, helps to clarify
the influence of species' interaction on mass extinctions.

The amount of newly introduced species per time-step should be
proportional to the number of already existing species, with some
constant of proportionality $g$ (the growth rate). This gives an
unbounded exponential growth, which is in good agreement with the data
of Benton~\cite{Benton}. However, since recourses on
earth are finite, the growth of the species must be
limited as well. Therefore, we believe it is justified to introduce a
logistic factor $(1-\frac{N}{N_{\max}})$, where $N_{\max}$ is the
maximal number of species that 
can be sustained with the available resources. With this factor it is
possible to work with a finite model. A few comments on the
fact that in nature this $N_{\max}$ is probably not constant in time will be
given later.

For the above reasons we want our system to grow according to
the differential equation
\begin{equation}\label{eq1}
  \frac{{\rm d}N}{{\rm d}t} = gN(1-\frac{N}{N_{\rm max}})\,.
\end{equation}
Since our model is discrete, in
time as well as in the number of species,
instead of (\ref{eq1}) we use the corresponding difference equation 
\begin{equation}\label{eq3}
  \Delta N(t+\Delta t)=
      \frac{N(t)N_{\rm max} e^{g\Delta t}}
            {N_{\rm max} +N(t)(e^{g\Delta t}-1)}-N(t)\,,
\end{equation}
where $\Delta t$ is one simulation time-step (usually set equal to 1).
As $\Delta N$ has to be an integer, we use the 
fractional part of $\Delta N$ as probability to round up or down.
In the
limit $g\rightarrow 0$ (which corresponds to $\Delta t\rightarrow 0$)
Equation~(\ref{eq3}) reduces to 
Equation~(\ref{eq1}). In the limit $g\rightarrow\infty$
Equation~(\ref{eq3}) becomes $\Delta N=N_{\rm max}-N$, which means
that our model reduces to the original one in the limit of an infinite
growth rate. 

Now we can formulate our model: We set $\Delta t=1$. At every time
step, a stress value $\eta$ is chosen and
all species with $x_i<\eta$ are removed. Then, an amount
$\Delta N$ of new species is introduced into the system. Finally, a
fraction $f$ of the species is assigned new thresholds.  

\begin{figure}
\narrowtext
\centerline{
        \epsfxsize=\columnwidth{\epsfbox{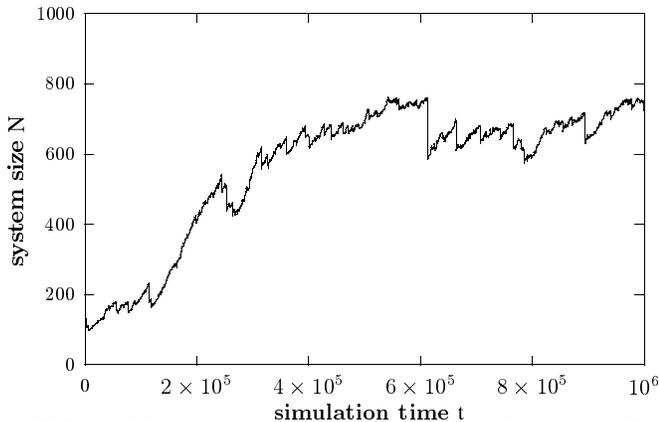}}
}
\caption{The evolution of the system size $N$ in time. The parameters
  are $g=4\times10^{-5}$, $\sigma=0.05$, $f=10^{-5}$, and $N_{\max}=1000$ with
  exponentially distributed stress.
\label{Fig1}}
\end{figure}

\begin{figure}
\narrowtext
\centerline{
        \epsfxsize=\columnwidth{\epsfbox{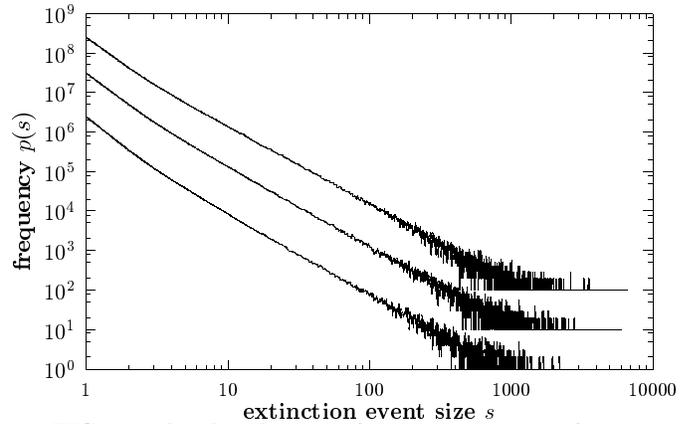}}
}
\caption{The distribution of extinction events for a system with
  exponentially distributed stress, $\sigma=0.05$ and 
  $N_{\max}=10000$. The growth rate is,
  from bottom to top, $g=4\times 10^{-5}$, $g=0.002$, $g=10$. It can
  be seen that the power-law behavior does depend only marginally on
  the growth 
  rate. The curves have been rescaled so as not to overlap.
\label{Fig2}}
\end{figure}

A typical evolution of the system size $N$ in time is presented in
Figure~\ref{Fig1}. The process of growth of new species is constantly
disrupted by small extinction events. From time to time, bigger
events, which disturb the system significantly, occur. A plot of the
distribution of extinction events (Figure~\ref{Fig2}) shows a
power-law decrease. Variation of the growth rate over several orders
of magnitude does change the exponent only slightly. 

We can explain the exponent of the power-law by extending the analysis
of Sneppen and Newman to our model.
The probability that species leave a small
intervall ${\rm d}x$ of the time averaged distribution $\bar\rho(x)$ is
proportional to $(f+p_{\rm move}(x))\bar\rho(x)$, where 
$p_{\rm move}(x)$ is the probability that a species with threshold $x$
is hit by stress. Let $\alpha$ be a variable that measures the
``emptiness'' of the system, i.e.\ $\alpha\propto 
(1-N/N_{\rm max})$. The rate at which the intervall ${\rm d}x$ is
repopulated is then proportional to $(f(1-\alpha)+
g\alpha(1-\alpha))p_{\rm thresh}(x)$ in the limit $\Delta
t\rightarrow 0$. In equilibrium the rates of species' loss and
repopulation balance, and we find the master-equation
\begin{equation}\label{eq4}
  (f + p_{\rm move}(x))\bar\rho(x)=
      (f(1-\bar\alpha)+ g\bar\alpha(1-\bar\alpha))p_{\rm thresh}(x)\,.
\end{equation}
Note that we had to replace $\alpha$ by its
time-averaged value $\bar\alpha$ and that we can always take the limit $\Delta
t\rightarrow 0$ in the steady-state.
After rearranging Equation~(\ref{eq4}), we find 
\begin{equation}\label{eq5}
  \bar\rho(x)=(f(1-\bar\alpha)+g\bar\alpha(1-\bar\alpha))
                \frac{p_{\rm thresh}(x)}
                 {f+p_{\rm move}(x)}\,.
\end{equation}
Equation (\ref{eq5}) can be solved if we choose how to normalize
$\bar\rho(x)$ and $\bar\alpha$. Since we can think of the system as
containing $N_{\rm max}$ species at any time step, from which there are $N$
active and $N_{\rm max}-N$ dead, it makes sense to normalize the sum
of $\bar\alpha$ and $\bar\rho(x)$ to unity, viz.
\begin{equation}
  1 = \bar\alpha + \int \bar\rho(x) {\rm d}x\,.
\end{equation}
That implies, nevertheless, that we do not normalize
$\bar\rho(x)$ to unity. Rather, 
$\int\bar\rho(x){\rm d}x$ gives the ratio $\bar N/N_{\rm max}$. 

For $\bar\alpha$ we find, apart from the trivial solution $\bar\alpha=1$, the
solution $\bar\alpha=(A-f)/g$, with
\begin{equation}
 A^{-1} =\int \frac{p_{\rm thresh}(x)}{f+ p_{\rm move}(x)} 
       {\rm d}x\,.
\end{equation}
For $\bar\rho(x)$, we find
\begin{equation}\label{eq6}
  \bar\rho(x)=A\left(1-\frac{A-f}{g}\right) \frac{p_{\rm thresh}(x)}
                 {f+p_{\rm move}(x)}\,.
\end{equation}
We thus have the interesting result that apart from the
overall factor $1-\bar\alpha$, which determines the average system size,
the shape of $\bar\rho(x)$ is identical to that found by Sneppen and
Newman.
Since only the shape $\bar\rho(x)$, but not the overall factor, is
responsible for the power-law distribution of extinction events (for
details see~\cite{Newman1}) we find that, within the time averaged
approximation, the exponent $\tau$ of the power-law decrease is exactly
the same as in the original model, even for very small $g$. 

If we take the limit $g\rightarrow\infty$ in Equation~(\ref{eq6}) we
can restore the 
expression found by Sneppen and 
Newman, which was to be expected
since our model reduces to the original one in that limit. In the
region of very small $g$, we can read off from Equation~(\ref{eq6})
that the system breaks down at a critical growth rate $g_c=A-f$. This
is the case when the growth rate is so small that the regrowth of
species cannot compensate the successive extinction events. Every
system with $g<g_c$ will eventually end up with $N=0$, regardless of
the number of species at the beginning of the simulation. 

\begin{figure}
\narrowtext
\centerline{
        \epsfxsize=\columnwidth{\epsfbox{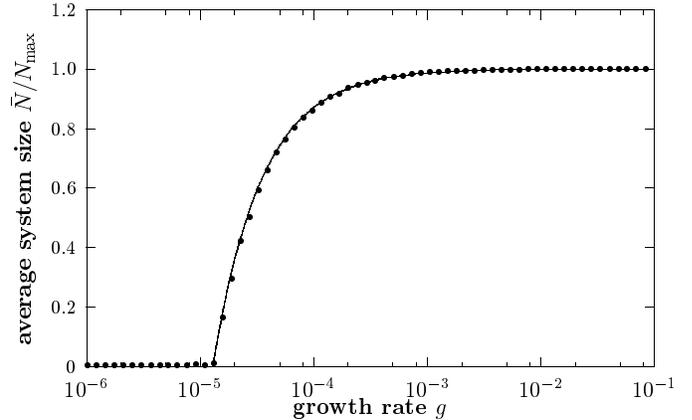}}
}
\caption{The average system size $\bar N$ versus the growth rate~$g$.
  We used exponentially distributed stress
  with $\sigma=0.05$ and $f=10^{-5}$. The solid line is the
  analytic expression, the points are the simulation results. 
\label{Fig3}}
\end{figure}

\begin{figure}
\narrowtext
\centerline{
        \epsfxsize=\columnwidth{\epsfbox{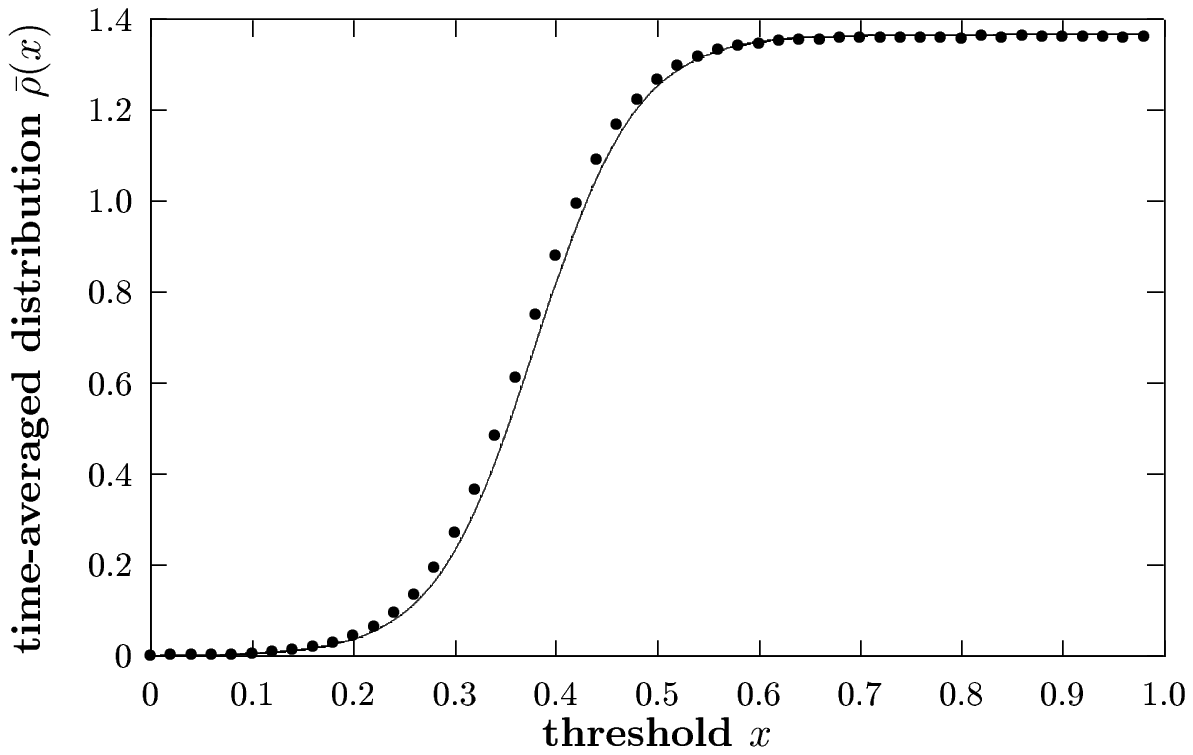}}
}
\caption{The time averaged distribution $\bar\rho(x)$. The parameters
  used are $g=0.002$, $\sigma=0.05$, and $f=5\times 10^{-4}$ with
  exponentially distributed stress. The solid line is the analytic
  expression, the points are the simulation results. 
\label{Fig4}}
\end{figure}

For the simulation results presented here we have used exponentially
distributed stress only, i.e., $p_{\rm
  stress}(\eta)=\exp(-\eta/\sigma)/\sigma$. We did simulations with
 $N_{\max}$ between 1000 and
10000. Figure~\ref{Fig3} shows 
the dependence of the average system size $\bar N$ of $g$. We can
clearly see the breakdown of the system at $g_c$. A measurement of the
time-averaged distribution of thresholds $\bar\rho(x)$ is presented in
Figure~\ref{Fig4}. The exponent $\tau$ of the power-law distribution of 
extinction events is found to be $\tau=1.9\pm 0.1$ for $g=10$,
$\tau=2.0\pm 0.1$ for $g=0.002$, $\tau=2.05\pm 0.1$ for
$g=4\times10^{-5}$ (for exponentially 
distributed stress, $\sigma=0.05$, $f=10^{-5}$, Figure~\ref{Fig2}).
The exponent decreases slightly with increasing $g$. For $g=10$,
we have already good agreement with the exponent found by Newman and
Sneppen~\cite{Newman1} for $g=\infty$, viz. $\tau=1.85\pm 0.03$.

\begin{figure}
\narrowtext
\centerline{
        \epsfxsize=\columnwidth{\epsfbox{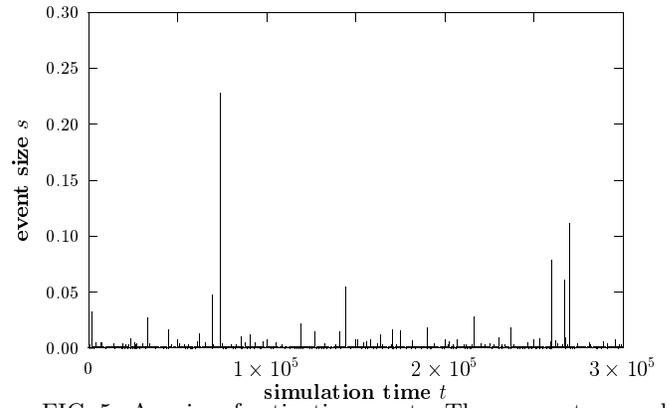}}
}
\caption{A series of extinction events. The parameters  
  used are $g=4\times10^{-5}$, $\sigma=0.05$, and $f=5\times 10^{-4}$ with
  exponentially distributed stress. Aftershocks cannot clearly be identified.
\label{Fig5}}
\end{figure}

An interesting feature of the original model by Newman and Sneppen is
the existence of aftershocks, a series of smaller events following a
large one.  These aftershocks have their origin in the fact that after
a large event the introduction of new species reduces significantly the
mean threshold value, thus increasing the probability to get
further events. Since the existence of aftershocks is a result of the
immediate refilling of the system after an event, we cannot
necessarily expect to 
see aftershocks when the refilling is done in finite time, especially
at a small growth rate. Numerical simulations show that there are
aftershocks for larger values of $g$, but when $g$ approaches $g_c$,
aftershocks cannot clearly be identified anymore. The region where
this happens is that in which the average system size
decreases rapidly with $g$. For these values of $g$, the typical time
the system needs to regrow the amount of species lost in a major event
exceeds the typical time needed to create a major stress value.
In Figure~\ref{Fig3}, the region in which we do not find aftershocks is
between $g=g_c=1.3\times 10^{-5}$ and about $g=5\times 10^{-4}$. A
typical example for a series of events in a system with $g$ close to
$g_c$ is presented in Figure~\ref{Fig5}. 

Sneppen and Newman argued that the
existence of aftershocks  might provide a measure to
distinguish between  coherent-noise driven systems and SOC
systems. This is certainly true in the sense that systems exhibiting
aftershocks are better candidates for coherent-noise driven systems
rather than for SOC systems. But our simulations show that there are
systems without clear aftershocks that still should be classified as
coherent-noise driven.

We have focused on logistic growth since we believe it is suitable
for the study of mass extinctions.
In principle it is possible to use different types of
growth. We have done some simulations with linear growth, where in every
time-step a fixed amount of new species is introduced into the system,
as long as $N<N_{\max}$. These simulations indicate that the
respective type of growth used does not affect
the appearance of a power-law distribution with exponent almost
independent from the growth rate.
But whether aftershocks appear or not, is
indeed dependend on the type of growth we choose. In a system with linear
growth aftershocks can be seen clearly even for small growth rates.

If we want to use a coherent noise model with variable system size as
a model of biological evolution, some remarks about the meaning of
$N_{\rm max}$ are necessary. The fact of allowing the regrowth of
species in finite time, instead of refilling the system immediately,
represents a first step closer to reality. But for ecosystems it is
certainly not a good
assumption to keep the maximal system size $N_{\rm max}$ fixed, since
the number of species an ecosystem can contain depends on the
interaction of species themselves. Therefore, a next step could be
to change $N_{\rm max}$ after every extinction, e.g., up or down by
chance and by an amount
proportional to the size of the event. This is motivated by the fact
that bigger events are expected to be correlated with a more
profound restructuring of the ecosystem, and as simulations show we
still find power-law distributions with exponents $\tau\approx2$. The
behaviour of 
such a system has a very rich structure with long times of relatively
little change (stasis) and sudden bursts of evolutionary activity
(punctuated equilibrium), where a major extinction event is followed by
a regrowth of species to a system size much bigger than the one before
the event. The so found curves of the system size $N$ agree
qualitatively well with the fossil record~\cite{Benton}.

We have generalized a coherent noise model to a model with variable
system size. The most important feature of coherent noise models, the
power-law distribution of event sizes with an exponent close to 2,
does not change under the generalization. This means that the validity of
Newman's approach to explain biological extinction with a coherent
noise model is not affected by the regrowth of
species in finite time. An interesting new feature that emerges from a
variable system size is the existence of a critical growth rate $g_c$.
Systems with $g<g_c$ will always end up with $N=0$ after some
time. Therefore in a world in which the regrowth of species is too slow
to compensate external influences no stable ecosystems can exist. In
the framework of our model we conclude that the process of mutation
and diversification of 
species at sufficiently high rate is necessary for the stability of
life on earth. 

We thank Stephan Altmeyer for stimulating discussions.



\end{multicols}


\begin{references}

\bibitem{Raup} D. M. Raup, Science {\bf 231}, 1528 (1986).
\bibitem{Bak1} P. Bak, C. Tang, and K. Wiesenfeld,
  Phys. Rev. Lett. {\bf 59}, 381 (1987).
\bibitem{Kauffman1} S. A. Kauffman and S. J. Johnsen, Theoretical
  Biology {\bf 149}, 467 (1991); S.~A.~Kauffman, {\em The Origins of
  Order}, OUP, Oxford 1992.
\bibitem{Bak2} P. Bak, K. Sneppen, Phys. Rev. Lett. {\bf 71}, 4083
  (1993).
\bibitem{Manrubia} S. C. Manrubia and M. Paczuski, cond-mat/9607066.
\bibitem{Newman2} M. E. J. Newman, Proc. R. Soc. Lond. B{\bf 263},
  1605 (1996);
  M. E. J. Newman, adap-org/9702003.
\bibitem{Newman1} M. E. J. Newman and K. Sneppen, Phys. Rev. E{\bf
  54}, 6226 (1996); K. Sneppen and M. E. J. Newman, cond-mat/9611229.
\bibitem{Benton} M. J. Benton, Science 268, 52 (1995).
\bibitem{Vandewalle} N. Vandewalle and M. Ausloos, Physica D{\bf 90},
  262 (1996).
\bibitem{Head} D. A. Head and G. J. Rodgers, Phys. Rev. E{\bf 55},
  3312 (1997).
\bibitem{Newman3}Sneppen and Newman mention a lattice-version of their
  model in cond-mat/9611229 that incorporates interaction and behaves
  very similar to the non-interacting version. Furthermore, Roberts
  and Newman have proposed a model that can be seen as a fusion
  between the Bak-Sneppen model and the coherent-noise dynamic. Their
  results (especially the distribution of extinction events) seem to
  be dominated by the coherent-noise dynamic (B. W. Roberts,
  M. E. J. Newman, J. Theor. Biol. {\bf 180}, 39 (1996)).


\end{references}
\end{document}